\documentclass{emulateapj}

\usepackage{amsmath}
\usepackage{amssymb}
\usepackage{graphicx}
\usepackage{color}
\usepackage{natbib}
\usepackage{epsfig}
\usepackage{placeins}
\usepackage{hyperref}
\usepackage{comment}
\graphicspath{ {figuresNEW/} }

\def\gtaprx {\lower ee.1ex\hbox{\rlap{\raise .6ex\hbox{\hskip .3ex
	{\ifmmode{\scriptscriptstyle >}\else
		{$\scriptscriptstyle >$}\fi}}}
	\kern -.4ex{\ifmmode{\scriptscriptstyle \sim}\else
		{$\scriptscriptstyle\sim$}\fi}}}
\def\ltaprx {\lower .1ex\hbox{\rlap{\raise .6ex\hbox{\hskip .3ex
	{\ifmmode{\scriptscriptstyle <}\else
		{$\scriptscriptstyle <$}\fi}}}
	\kern -.4ex{\ifmmode{\scriptscriptstyle \sim}\else
		{$\scriptscriptstyle\sim$}\fi}}}

\newcommand{\cutt}[1]{\textcolor{blue}{}}

\newcommand{\Ms}{{\ensuremath{M_{\odot} }}}

\newcommand{\C}{{\ensuremath{^{12}\mathrm{C}}}}

\newcommand{\N}{{\ensuremath{^{14} \mathrm{N}}} }

\begin{document}

\title{Finding Lensed Direct-Collapse Black Holes and Supermassive Primordial Stars}

\author{Anton Vikaeus\altaffilmark{1}, Daniel J. Whalen\altaffilmark{2} and Erik Zackrisson\altaffilmark{1}}

\altaffiltext{1}{Observational Astrophysics, Department of Physics and Astronomy, Uppsala University, Box 516, SE-751 20 Uppsala, Sweden} 

\altaffiltext{2}{Institute of Cosmology and Gravitation, Portsmouth University, Dennis Sciama Building, Portsmouth PO1 3FX}

\begin{abstract}

Direct-collapse black holes (DCBHs) may be the seeds of the first quasars, over 200 of which have now been detected at $z >$ 6.  The {\em James Webb Space Telescope} ({\em JWST}) could detect DCBHs in the near infrared (NIR) at $z\lesssim20$ and probe the evolution of primordial quasars at their earliest stages, but only in narrow fields that may not capture many of them.  Wide-field NIR surveys by {\em Euclid} and the {\em Nancy Grace Roman Survey Telescope} ({\em RST}) would enclose far greater numbers of DCBHs but only directly detect them at $z \lesssim$ 6 - 8 because of their lower sensitivities.  However, their large survey areas will cover thousands of galaxy clusters and massive galaxies that could gravitationally lense flux from DCBHs, boosting them above current {\em Euclid} and {\em RST} detection limits and revealing more of them than could otherwise be detected.  Here, we estimate the minimum number density of strongly lensed DCBHs and supermassive primordial stars required for detection in surveys by {\em Euclid}, {\em RST} and \textit{JWST} at $z \lesssim$ 20.  We find that for reasonable estimates of host halo numbers {\em RST}, {\em Euclid}, and {\em JWST} could potentially find hundreds of strongly-lensed DCBHs at $z =$ 7 - 20. \textit{RST} would detect the most objects at $z\lesssim10$ and \textit{JWST} would find the most at higher redshifts. Lensed supermassive primordial stars could potentially also be found, but in fewer numbers because of their short lifetimes.

\end{abstract}


\maketitle

\section{Introduction}

Direct-collapse black holes could be the seeds of the first quasars, nine of which have now been found at $z > 7$ \citep{mort11,ban18,wang21}.  They are thought to form when primordial haloes in unusual environments grow to masses of $\sim$ 10$^7$ \Ms\ \citep{agarw15,latif15a,hir17} and collapse via atomic cooling at rates of up to $\sim$ 1 \Ms\ yr$^{-1}$ \citep[e.g.,][]{rd18,pat21a}.  Stellar evolution models indicate that collapse leads to supermassive stars (SMSs) that reach a few 10$^4$ - 10$^5$ \Ms\ before collapsing to DCBHs  \citep{hos13,tyr17,hle18a,hle18b,tyr21a,herr21a}.  DCBHs are thought to be the seeds of the first quasars because it is difficult for normal Pop III star BHs to grow rapidly after birth \citep{wan04,wf12,srd18}.  They are born with much larger masses and in much higher densities in halos that retain their fuel supply even when it is heated by X-rays (\citealt{jet13} -- see \citealt{titans} for recent reviews).  A DCBH must lie at the nexus of cold accretion flows \citep[e.g.,][]{dm12} or merge with other gas-rich halos capable of fueling its rapid growth \citep[e.g.,][]{li07} to become a SMBH at $z >$ 6.  Radiation hydrodynamical simulations show that 10$^5$ \Ms\ DCBHs at $z \sim 19$ can grow to $2 \times 10^9$ \Ms\ by $z =$ 7.1 in cold accretion flows \citep{smidt18,latif20b,zhu20,vgf21} and that these flows  can form DCBHs without the need for exotic environments or even atomic cooling \citep{latif22b}.

\begin{figure*} 
\begin{center}
\includegraphics[width=\columnwidth]{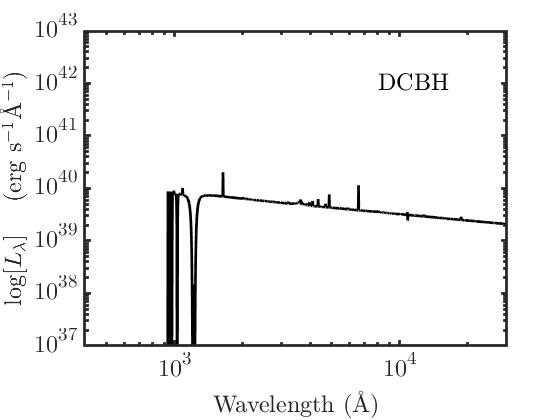} 
\includegraphics[width=\columnwidth]{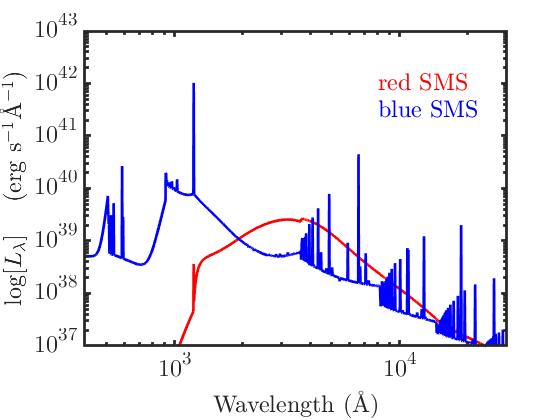}  
\end{center}
\caption{Source frame spectra for the 10$^5$ \Ms\ DCBH (left) and 1.0 \Ms\ yr$^{-1}$ SMSs (right) after reprocessing by the host gas envelope.}
\vspace{0.1in}
\label{fig:spectra} 
\end{figure*}

DCBHs could be found at $z \sim$ 8 - 10 by the Square Kilometer Array and next-generation Very Large Array \citep{wet20a}, at $z \sim$ 10 by future X-ray missions such as the {\em Advanced Telescope for High-Energy Astrophysics} ({\em ATHENA}) and {\em Lynx}, and at $z \sim$ 20 by the {\em James Webb Space Telescope} \citep[{\em JWST};][]{pac15,2016Pacucci,nat17,bar18,wet20b}.  NIR from their SMS progenitors could be detected at $z \sim$ 10 by {\em JWST} \citep{sur18a,sur19a} and gravitational waves from DCBH mergers might be found by the Laser Interferometer Space Antenna \citep[LISA;][]{til18a,latif20a}.  

{\em JWST} could detect primordial quasars at their earliest stages of evolution but only in narrow fields of view that may not capture many of them. Wide-field surveys by {\em Euclid} and {\em RST} could enclose far greater numbers of DCBHs but only detect them at $z \lesssim$ 6 - 8 because of their lower sensitivities \citep{wet20b}. Furthermore, because the filters in {\em Euclid} and {\em RST} are limited to $\lambda < 2 \,\mu$m and $\lambda < 2.3 \,\mu$m, they cannot detect SMSs or DCBHs at $z \gtrsim$ 15 and $z \gtrsim 18$ respectively, because flux blueward of the corresponding wavelength in the rest frame of the object is resonantly scattered and absorbed by the neutral intergalactic medium (IGM) prior to the end of cosmological reionization. 

However, {\em Euclid} and {\em RST} could still find DCBHs or SMSs at $z \gtrsim$ 10 because their wide-field surveys (several tens or thousands of square degrees, respectively) will enclose thousands of galaxy clusters and massive galaxies that could gravitationally lense flux from these primordial sources. Strong lensing could offset the lower sensitivities of wide-area surveys and reveal more objects than could otherwise be detected (and at higher redshifts). Until now, surveys have targeted individual well-resolved cluster lenses with extreme magnifications, $\mu$, exceeding $\sim$ 1000 \citep{2022Welch} to search for faint, high-redshift objects. Lensed galaxies at $z \gtrsim 6$ have already been discovered at high magnifications (e.g., \citealt{brd14,vanz14,2020Salmon}). Wide-field surveys could produce more detections of high-$z$ objects because of their much larger total lensing areas, even though their magnifications are more modest \citep[e.g.,][]{vik21a}.  Here, we estimate the number of strongly-lensed DCBHs and SMSs that could be found in wide-field surveys by {\em Euclid} and {\em RST} and in deep {\em JWST} surveys. In Section 2 we discuss our DCBH and SMS source-frame spectra and statistical model for strong lensing in wide fields.  We show SMS and DCBH number counts in {\em Euclid}, {\em RST} and {\em JWST} surveys in Section 3 and conclude in Section 4.

\section{Numerical Method}

We convolve rest frame spectra for DCBHs and red and blue SMSs with statistical estimates of strong lensing in wide or narrow fields to calculate the minimum number density of objects required for just one object to appear in a given survey over a range of redshifts.  

\subsection{DCBH / SMS Spectra}

Studies of SMS evolution show that they either follow cool red tracks at temperatures of $\sim$ 10$^4$ K if they grow at the Hayashi limit or hot, blue tracks with temperatures above 10$^5$ K.  We take spectra for 10$^5$ \Ms\ DCBHs and 10$^5$ \Ms\ red and blue SMSs growing at 1 \Ms\ yr$^{-1}$ from \citet{wet20b} and \citet{sur18a,sur19a} respectively, which are shown in Figure~\ref{fig:spectra}.  Reprocessing of radiation from the object by the dense cocoon of atomically-cooled gas in which it forms was included in the spectrum calculation, and absorption by the atmosphere of the blue SMS was included in its spectrum.  Absorption by the envelope of the DCBH is evident from the attenuation of its flux blueward of the Lyman limit and the emission features at longer wavelengths.  Similar features are visible in the spectra of both stars.  Absorption and emission features are more prominent in the spectrum of the blue star because of its ionizing UV flux and absorption by its atmosphere.  The spectral limit of the red SMS occurs at a longer wavelength than for the SMS because of its lower temperature, but more flux emerges from the red star at wavelengths above 3000 \AA, which has important consequences for its detectability at higher redshifts.  Likewise, the flatter spectrum of the DCBH makes it easier to detect than SMSs at most redshifts, as will be shown later.

\begin{table}
\centering
\caption{Filters, areas and photometric depths for proposed \textit{JWST} (NIRCam/MIRI), \textit{Euclid}, and \textit{RST} surveys.}
\begin{tabular}{lcc} 
\hline 
								  \\
Filter & Area (deg$^2$) &  Depth (AB mag) \\
\ & \ & 5$\sigma$ point-source \\
\hline
							      \\
deep NIRCam: & \ &\ \\
F200W/F356W/F444W &  0.013  &  30.6/30.1/29.8   \\
med. NIRCam: & \ &\ \\
F200W/F356W/F444W & 0.053  &  29.7/29.3/29.0   \\
deep MIRI (F770W) & 0.0022 & 27.4      \\
med. MIRI (F770W) & 0.0039 & 25.6      \\
wide \textit{RST} (F184/F213) & 2000 & 26.3/25   \\
wide \textit{Euclid} (H) & 15000 & 24       \\
							      \\
\hline
                                                                          \\
ultra deep NIRCam: & \ &\ \\
F200W/F356W/F444W &  0.0027 &  31.8/31.3/31.0   \\
ultra deep MIRI (F770W)       & 0.0006 & 27.8  \\
deep \textit{RST}  (F184/F213)       & 40        & 28.5/27.2  \\
deep \textit{Euclid} (H)           & 40        & 26     \\
                                                                           \\
\hline  \\
\end{tabular}
\label{tbl:param}
\end{table}

\begin{figure}
\includegraphics[width=\columnwidth]{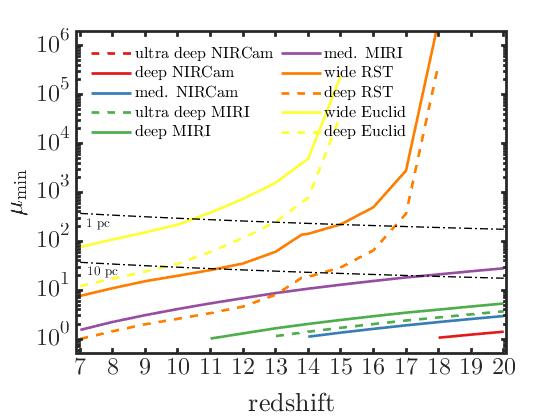}
\includegraphics[width=\columnwidth]{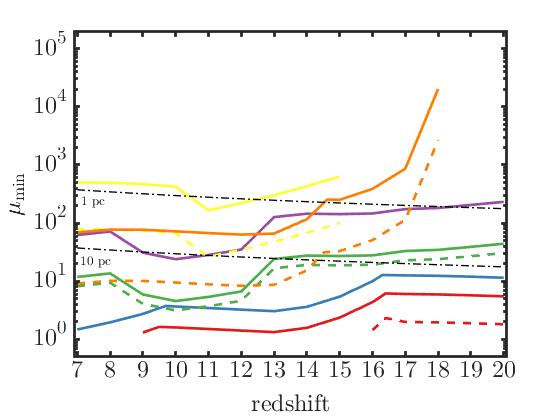}
\includegraphics[width=\columnwidth]{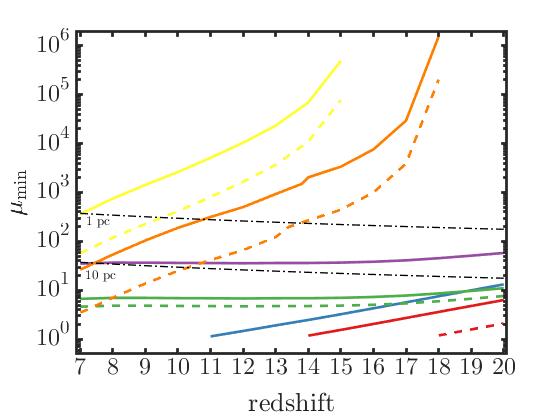}
\caption{Minimum magnification ($\mu_\mathrm{min}$) required for photometric detection of a gravitationally lensed DCBH (top), blue SMS (center) and red SMS (bottom) as a function of redshift for the different surveys. The filter used for each object and survey is specified in table \ref{tbl:zrange}. Note that the lines are only plotted for $\mu > 1$. Therefore the ultradeep NIRCam exposures are not included for the DCBH because no magnification is required in that particular case. The two dash-dotted black lines indicate the magnification above which an object with a given intrinsic size (1 pc and 10 pc plotted here) could become resolved in imaging.}  \label{fig:mag}
\vspace{0.1in}
\end{figure}

\subsection{Minimum Number Densities}

The minimum comoving number density ($n_\mathrm{min}$) of DCBHs or SMSs required to detect just one object in a randomly-oriented sky survey is 
\begin{equation} \label{eq:nmin}
    n_\mathrm{min} = \frac{1}{V_c(A,z,\Delta z)\times P(\geq \mu_\mathrm{min},z)} \frac{\Delta t(z,\Delta z)}{\tau},
\end{equation}
where $P(\geq \mu_\mathrm{min},z)$ is the probability of achieving the minimum magnification required to boost flux from the object above the detection threshold for the survey and is calculated from the lensing model of \citet[][see Figure 1 therein for plots of $P(\geq \mu,z)$ for $\mu \ge$ 10, 100 and 1000]{zak15}. $V_c$ is the comoving volume spanned by the survey area $A$ of the telescope at a redshift $z$ with a redshift bin $\Delta z$ corresponding to those used in dropout selection techniques when observing with photometric filters (here, we set $\Delta z =$ 1). The effect of $\Delta z$ on $n_\mathrm{min}$ is partly compensated by the ratio of the cosmic time spanned by the survey volume, $\Delta t$, and the time for which the object remains uniquely identifiable as a DCBH or SMS, defined here as $\tau$.  For larger choices of $\Delta z$, the corresponding cosmic time $\Delta t$ is also greater, mitigating the dependence of $n_\mathrm{min}$ on $\Delta z$. We assume $\tau =$ 40 Myr for the DCBH, about one Salpeter e-folding time \citep{sal64}, after which its luminosity would begin to change significantly. This assumption is somewhat conservative as the luminosity is expected to increase over time. The SMSs have $\tau =$ 200,000 yr because they are much shorter lived. 

Our lensing model uses the large scale dark matter distribution of the Millennium simulation with a semi-analytical model \citep{2005Baugh} for the distribution of galaxies inside dark matter halos. From this distribution the lensing potential for a sequence of lens planes along different lines of sight can be calculated. Magnifications for randomly oriented sightlines are then calculated as a function of redshift for the source to produce the magnification probability function $P(\geq \mu,z)$. \citet{2020Meneghetti} argued that failing to consider substructures in galaxy clusters could lead to underestimates of the likelihood of strong gravitational lensing. Since our dark matter distributions do not resolve these substructures, our estimates of DCBH and SMS detections should be considered to be conservative.

As constructed in Equation~\ref{eq:nmin}, $n_{min}$ establishes the requirements on minimum object number density for a 1$\sigma$ detection such that any randomly-oriented sky survey will find at least one object in the survey volume with a probability of $P_\mathrm{det} \sim 0.68$. This constraint is set by $P_\mathrm{det} = 1-[1-P(> \mu\mathrm{_{min}},z)]^{N}$, where $N$ is the number of SMSs or DCBHs within the survey area with an age that does not exceed their respective $\tau$. Requiring a specific probability $P_\mathrm{det}$ therefore sets a relation between $N$ and $P(\geq \mu,z)$ that can be varied by requiring, for example, a 2$\sigma$ detection ($P_\mathrm{det} \sim 0.95$). This increases the required number of objects inside the survey volume by a factor of 3 compared to the 1$\sigma$ detection for any given $P(\geq \mu,z)$. Here we consider 1$\sigma$ detections to set a lower limit on $n_{min}$.

\subsection{Resolved/Unresolved Imaging}

The DCBH would appear as a point source given the magnifications necessary to detect it in the surveys here because we assume its spectrum is dominated by its accretion disk and that nebular emission is negligible. The intrinsic size of the DCBH is then tens of AU, not parsecs, which could be the case for the nebula. The top panel of Figure~\ref{fig:mag} shows the magnification required to detect a 10$^5$ \Ms\ DCBH. The black dash-dotted lines indicate the magnifications above which an object with an intrinsic size of 1 pc or 10 pc could become resolved in imaging. Here, we make the assumption that the stretching of the image due to lensing is only in one direction. In reality the object will also be stretched by a non-negligible amount in other directions, which would allow for higher magnifications, making this a conservative estimate. If there is significant nebular emission, Figure~\ref{fig:mag} shows that, e.g., the deep \textit{RST} survey would only detect DCBHs that are resolved in imaging at $z\gtrsim 14.5$ if they are 10 pc sources or $z\gtrsim 16.5$ if they are 1 pc sources. Note that a DCBH or SMS with an intrinsic size of $\sim 10$ AU would remain point-like (unresolved) in all surveys since $\mu \gtrsim 4\times 10^6$ would be required throughout the redshift range to resolve such a small object.

Similar arguments apply to SMSs, which can also be taken to be point sources if their intrinsic size is on the order of AU.  Nebular emission in blue SMS spectra can produce a larger object that could be resolved at higher magnifications depending on the size of the nebula (center and bottom panels of Figure~\ref{fig:mag}).  If DCBHs and SMSs are indeed tens of AU in size, their point-like appearance even at high magnifications enables their detection at very high redshifts.  We find that \textit{JWST} NIRCam imaging never requires significant magnifications and would therefore be able to detect DCBHs and SMSs as point-like sources throughout the redshift range, even for objects with intrinsic sizes on the order of parsecs. When objects become resolved, additional modelling of the sources on smaller scales is required to determine if the detection is a DCBH or an SMS.

\begin{table}
\centering
\caption{Photometric filters providing the lowest minimum number densities for the different surveys and their corresponding redshift ranges. \textit{JWST} filters are selected from the surveys in \citet{riek19}.}
\begin{tabular}{llll}
\hline 
						                                    \\
Object & Survey & Filter & Redshifts                             \\
\\
\hline
\\
DCBH  &  {\em Euclid}  & H           &   $z =$ 7 - 15        \\
DCBH  &  {\em RST}     & F184     &   $z =$ 7 - 13.8     \\
DCBH  &  {\em RST}     & F213     &   $z =$ 13.8 - 18   \\
DCBH  &  NIRcam         & F444W &   $z =$ 7 - 20        \\
DCBH  &  MIRI              & F770W  &   $z =$ 7 - 20        \\
\\
\hline
\\
BSMS  &  {\em Euclid}  & H           &   $z =$ 7 - 15        \\ 
BSMS  &  {\em RST}     & F184     &   $z =$ 7 - 14.5     \\
BSMS  &  {\em RST}     & F213     &   $z =$ 14.5 - 18   \\
BSMS  &  NIRcam UD  & F200W  &    $z =$ 7 - 16.4     \\
BSMS  &  NIRcam UD  & F356W  &    $z =$ 16.4 - 20   \\
BSMS  &  NIRcam D     & F356W  &    $z =$ 7 - 9.5       \\
BSMS  &  NIRcam D     & F200W  &    $z =$ 9.5 - 16.4  \\
BSMS  &  NIRcam D     & F356W  &    $z =$16.4 - 20    \\
BSMS  &  NIRcam M     & F356W  &    $z =$ 7 - 9.7       \\
BSMS  &  NIRcam M     & F200W  &    $z =$ 9.7 - 16.3  \\
BSMS  &  NIRcam M     & F356W  &    $z =$ 16.3 - 20   \\
BSMS  &  MIRI               & F770W  &   $z =$ 7 - 20        \\
\\
\hline
\\
RSMS  &  {\em Euclid}  & H           &   $z =$ 7 - 15        \\
RSMS  &  {\em RST}     & F184      &    $z =$ 7 - 13.4     \\
RSMS  &  {\em RST}     & F213      &    $z =$ 13.4 - 18   \\
RSMS  &  NIRcam         & F444W  &    $z =$ 7 - 20        \\ 
RSMS  &  MIRI              & F770W  &   $z =$ 7 - 20        \\
\\
\hline
\multicolumn{4}{l}{\footnotesize Note: UD = ultra deep, D = deep, M = medium,} \\
\multicolumn{4}{l}{\footnotesize BSMS = blue SMS, RSMS = red SMS.}
\end{tabular}
\label{tbl:zrange}
\end{table}

\begin{figure}
\includegraphics[width=\columnwidth]{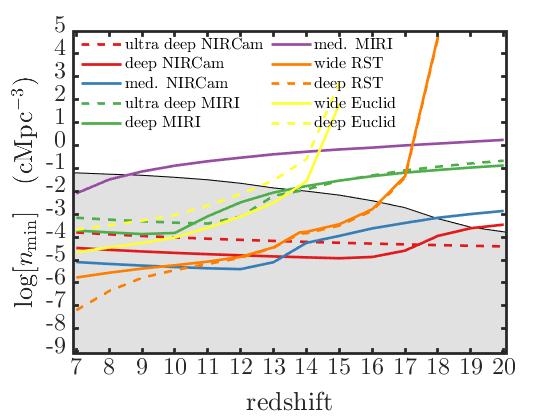}
\includegraphics[width=\columnwidth]{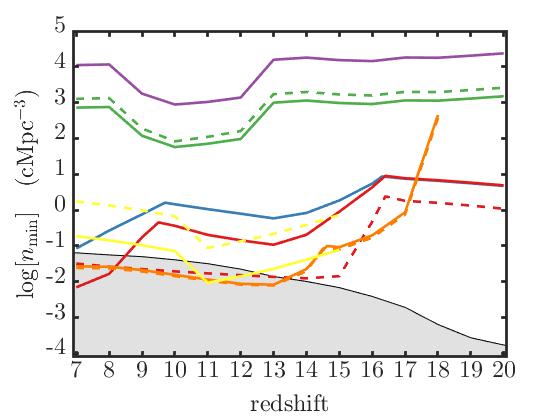}
\includegraphics[width=\columnwidth]{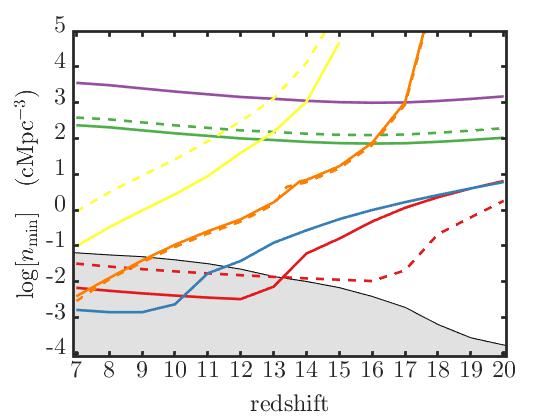}
\caption{Minimum number densities $n_\mathrm{min}$ required for photometric detection of one single gravitationally lensed DCBH (top), blue SMS (center), and red SMS (bottom) as a function of redshift. The gray shaded area (extending down to $10^{-9}$ Mpc$^{-3}$ in all figures) shows the expected range of number densities of halos that can host SMSs and DCBHs from simulations \citep{hab16,rosa17}.}  \label{fig:nmin}
\end{figure}

\subsection{Survey Parameters}

{\em JWST} NIRcam and MIRI filters range from 0.6 - 5 $\mu$m and 5.6 - 25.5 $\mu$m and {\em Euclid} covers 0.9 - 2 $\mu$m and {\em RST} covers 0.5 - 2.3 $\mu$m. As noted earlier, these filters allow {\em JWST} in principle to detect SMSs and DCBHs out to $z \sim$ 20 but restrict detections by {\em Euclid} and {\em RST} to at most $z \sim$ 15 and $z \sim$ 18 respectively.  Point-source $5\sigma$ photometric detection thresholds in proposed {\em JWST}, {\em RST} and {\em Euclid} surveys and their filters are shown in Table~\ref{tbl:param} \citep{riek19,2019Akeson,euclid}.  To maximize detections over the range of redshifts in our study, we consider two filters for {\em RST} since an object at higher redshifts (e.g., at $z\gtrsim 13.8$ for DCBHs) may be brighter in the F213 filter than the F184 filter (at $z\gtrsim 15$ the flux entering the F184 filter originates blueward of Ly$\alpha$ and is thus attenuated by the neutral IGM). We likewise consider four \textit{JWST} filters to optimize detections over specific ranges in redshift. The redshift ranges for which the filters in each survey were used to minimize $n_\mathrm{min}$ for each type of object are summarized in Table~\ref{tbl:zrange}. Note that in most of the cases the redshift ranges for the optimum filter does not depend on the depth of the survey.  In this paper we consider the {\em JWST} surveys summarized in \citet{riek19}, in which the filters that are available for the designated survey depth and areal coverage are limited. Because of the photometric depth of \textit{JWST}, DCBHs can be detected in several of its filters without gravitational lensing up to very high redshifts in ultra deep fields. When gravitational lensing is not required, the survey area alone determines the minimum required number density. 

\section{Results}

\subsection{DCBHs}
We show the minimum required number density ($n_\mathrm{min}$) for one single detection from $z =$ 7 - 20 for a 10$^5$ \Ms\ DCBH in the top panel of Figure~\ref{fig:nmin}.  The deep field survey with {\em RST} will detect more strongly-lensed DCBHs than its wide survey at $z\lesssim 12$, above which the wide {\em RST} survey will perform equally well. The wide {\em Euclid} survey will detect more DCBHs than its deep survey at all redshifts. The deep {\em RST} survey performs better than all other surveys at $z \lesssim 10.5$. However, {\em JWST} medium NIRcam imaging would find the most DCBHs at $10.5 \lesssim z \lesssim 13.2$ and deep NIRCam imaging would detect the most at $13.2 \lesssim z \lesssim 17.5$.  Single, ultra deep NIRCam pointings with 100 hr exposures will detect the most DCBHs at $z\gtrsim 17.5$.  We note that $n_\mathrm{min}$ for ultra deep NIRCam imaging varies with redshift only with $V_c$ and $\Delta t$ because DCBHs are visible in these exposures without gravitational lensing.  In contrast, lensing is required for all detections by wide and deep surveys with {\em RST} and {\em Euclid}, even at lower redshifts.  In these surveys, $n_\mathrm{min}$ rises rapidly at $z \sim$ 14 - 15 because of absorption of DCBH flux at corresponding wavelengths ($\lambda \leq \lambda_\mathrm{Ly\alpha}$) in the source frame by the neutral IGM, as discussed earlier. Although the NIRcam plots in Figure~\ref{fig:nmin} are for the F444W filter, the other NIRCam filters yield similar numbers of detections because the DCBH spectrum is relatively flat (see Figure~\ref{fig:spectra}).

We overlay the range of number densities of halos expected from simulations (defined here as $n_\mathrm{sim}$) to host DCBHs as the gray regions in Figure~\ref{fig:nmin} \citep{hab16,rosa17}.  They vary by many orders of magnitude with redshift because of uncertainties in Lyman-Werner and UV self-shielding in the halos but have a lower limit of $\sim$ $10^{-9}$ Mpc$^{-3}$ that is set by the number of observed quasars at $z\sim 6$.  For none of the surveys to detect a DCBH the number density of host halos would have to be less than $\sim 10^{-7}$  Mpc$^{-3}$ at $z \sim 7$ and $\sim 10^{-4.5}$ Mpc$^{-3}$ at $z\sim 20$, which are fairly severe limits. As $n_\mathrm{min}$ sets the required minimum number density for one single detection, the ratio $n_\mathrm{sim}/n_\mathrm{min}$ yields the number of objects expected to be detected at a given redshift. In the optimistic limit this indicates that these surveys could detect several hundred to several thousand lensed DCBHs per unit redshift at lower $z$ and up to ten even at the highest redshifts -- but these predictions are very uncertain due to the large range of $n_\mathrm{sim}$ at any given redshift. Note however that the regions marked by gray in Figure~\ref{fig:nmin} may somewhat underestimate the number of DCBHs expected to be present at a given redshift because the most recent simulations of the formation of atomically-cooled halos have shown that 1 - 5 DCBHs can form per halo \citep[e.g.,][]{pat21a}.

\subsection{SMSs}

The $n_\mathrm{min}$ for the blue and red SMSs are shown as a function of redshift in the center (blue SMS) and bottom (red SMS) panel of Figure~\ref{fig:nmin}. SMSs are intrinsically brighter than the DBCH at some redshifts in some of the filters, therefore one might naively expect that their $n_\mathrm{min}$ could be less than those for DCBHs.  However, their shorter lifetimes result in minimum required number densities that are orders of magnitude greater than those for DCBHs and would be detected at most to $z \sim$ 14, even by ultradeep NIRcam exposures. The breaks in some of the lines in the center panel of Figure~\ref{fig:nmin} are due to switching to the optimum filter at those redshifts and to significant variations in the brightness of the blue SMS spectrum at certain wavelengths. The troughs in the MIRI plots at $z \sim$ 8 - 13 in the center panel are due to strong Balmer emission lines from the blue star. The MIRI F770W filter probes the rest frame spectrum between $5000 - 9000$ \AA which contains these lines. This brightens the star and enhances the prospects for detections over this redshift range by MIRI.

If we take into account the numbers of SMS host halos per redshift predicted by simulations in gray in Figure~\ref{fig:nmin}, deep NIRcam imaging would detect the most blue SMSs at $z =$ 7 - 8 and wide {\em RST} exposures would find the most at $z =$ 8 - 13.5, followed closely by ultradeep NIRcam pointings.  The wide {\em Euclid} survey is the only other one that would detect blue lensed stars, at $z =$ 10.5 - 12.5.  Medium NIRcam imaging would find the most red SMSs at $z =$ 7 - 10.2 and deep NIRcam exposures would find the most at $z =$ 10.2 - 13.5.  Over both ranges in redshift F444W is the optimum NIRCam filter. We again note that the regions in gray in Figure~\ref{fig:nmin} somewhat underestimate the number of SMSs expected to exist per unit redshift because more than one star can form per halo. Unlike DCBHs, we find that SMSs generally must be gravitationally lensed at higher redshifts in order to be detected, even by ultradeep NIRCam imaging (see fig.~\ref{fig:mag}).

\section{Conclusion}

Our calculations show that {\em RST}, {\em Euclid}, and {\em JWST} would detect far more lensed DCBHs than lensed SMSs, primarily because the stars have such short lifetimes.  Indeed, the best way to constrain SMS numbers at high redshifts is to detect their DCBHs since SMS detections could effectively undercount their true numbers by orders of magnitude. Considering the surveys included in this paper, {\em RST} will detect the most lensed DBCHs for $z \lesssim 10$ and \textit{JWST} will find the most at $z \sim$ 10 - 20.  These surveys will find up to a few tens more red SMSs than blue ones because more of their flux is redshifted into the NIR today.  In principle, {\em RST}, {\em Euclid}, and {\em JWST} may find no DCBHs if the number density of source halos is small enough at high redshifts.  However, the failure to detect these objects would impose severe constraints on DCBH number densities, less than 10$^{-6}$ Mpc$^{-3}$ from $z =$ 7 - 20.  For reasonable estimates of progenitor halo numbers, {\em RST}, {\em Euclid}, and {\em JWST} could find hundreds or even thousands of strongly lensed DCBHs at $z =$ 7 - 20 in the coming decade.

\FloatBarrier
\acknowledgments
AV and EZ acknowledge funding from the Swedish National Space Board.

\bibliographystyle{apj}
\bibliography{refs}

\end{document}